# How Beaufort, Neumann and Gates met? Subject integration with spreadsheeting

Maria Csernoch, Julia Csernoch

University of Debrecen, Faculty of Informatics

csernoch.maria@inf.unideb.hu, csernoch.julia@inf.unideb.hu

**ABSTRACT**

Computational thinking should be the fourth fundamental skill, along with reading, writing, and arithmetic (3R). To reach the level where computational thinking skills, especially digital problem solving have their own schemata, there is a long way to go. In the present paper, a novel approach is detailed to support subject integration and building digital schemata, on the well-known Beaufort scale. The conversion of a traditional, paper-based problem and a data retrieval process are presented within the frame of a Grade 8 action research study. It is found that both students' content knowledge and their digital skills developed more efficiently than in traditional course book and decontextualized digital environments. Furthermore, the method presented here can be adapted to any paper-based problems whose solutions would be more effective in a digital environment and which offer various forms for building schemata both in the subject matter and informatics.

## 1. INTRODUCTION

### 1.1. Digitalization and digitization

According to Prensky [Prensky, 2001a] [Prensky, 2001b], digital students are overtaking the world. Among others, he also claims that digital natives are born with computers, consequently (1) they can develop on their own and they must do it on their own, since (2) education seems to be way behind them. However, research found proof that students do not have skills and abilities assigned to them by Prensky [Kirschner & De Bruyckere, 2017] [Csernoch et al., 2015], and developing their computational thinking skills [Wing, 2006] is a rather demanding and challenging task of education both from the side of students and teachers [Kirschner et al., 2006] [Lister, 2008]. The primary problem is how to engage students in solving real world problems, how they can find ways to meaningful digital solutions [Soloway, 1993], how they can control tools, instead of tools controlling them [Soloway, 1993] [Baranyi & Gilányi, 2013] [Baranyi et al., 2015] [Wolfram, 2020] [Wolfram, 2015]. On the other hand, it is also proven that education is way behind digitalization and digitization, and teachers are not prepared to fulfil their role in this novel era. All the aspects of TPCK (Technological Pedagogical Content Knowledge) [Angeli & Valanides, 2015] [Kadijevich et al., 2013] should be present in the teaching-learning process, but it is not so. In general, the problem we are faced with is complex, and it is high time to take some small steps. In our point of view, we must begin with finding the balance between classical and digital problem solving.

### 1.2. EDUCATION PRODUCTION SYSTEMS

In general, digital education is a push production system [Ohno, 1988] where tools, software interfaces, their novel features are in the focus, in the hope to build up huge knowledge inventory. However, the erroneous end-user products [Ben-Ari, 1999] [Ben-Ari & Yeshno, 2006] [Tufte, 2003] [Csernoch et al., 2022] [Csernoch et al., 2023], including spreadsheets [EuSpRIG, 2023] [Csernoch et al., 2015] [Csernoch et al., 2021] [Nagy et al., 2021], prove that this method cannot be either effective or efficient. Human mind is responsible for this defect, since it does not work as a commodity inventory; knowledge, especially unconnected pieces of data cannot be stored in the long-term memory [Kirschner et al., 2006] [Sweller et al., 2011] [Low et al., 2011].

It is well known in the EuSpRIG community that time spent on fabricating documents, browsing, and clicking aimlessly use up an enormous amount of both human and machine resources. On the other hand, erroneous documents and their modification generate further waste that has serious financial consequences. We are convinced, in accordance with Ohno that the key is to give human intelligence to the machine (both hardware and software), and the transition from the single to the multi-skilled operators (end-users) [Ohno, 1988].

The present work is part of a larger campaign to introduce a digital education pull production system in accordance with the Toyota or Lean Production System (TPS) [Ohno, 1988] [Krafcik, 1988] [Womack & Jones, 2003] [Modig & Åhlström, 2018]. In this special education system, the values are to improve teachers' and students' computational thinking skills and problem solving abilities, furthermore to eliminate waste. The principles are just-in-time and autonomation which are supported by unorthodox methods and tools and activities. The just-in-time principle means that, instead of teaching software interfaces and features, problems are selected to solve, and only a limited set of tools is taught; those pieces which are necessary to solve the problem. This set includes data analysis, algorithms known from programming, and integration of software tools (the combination of various programs). Autonomation (automation with a human touch) is supported by the functional programing approach of Sprego [Csernoch, 2014] [Csernoch & Biró, 2015] [Csapó et al., 2019] [Csapó et al., 2020] and requires human intervention when the analysis and the debugging of the output reveal any discrepancy from standard.

The problem presented in the paper, selected from our depository, aims to demonstrate how this digital educational pull system works in classes.

### 1.3. Previous results

It is without question that there is high number of valuable papers and studies which details methods and results on developing students' computational thinking skills with spreadsheeting. However, it is beyond the scope of the present paper to cover thoroughly these sources, for at least two reasons. (1) The subject is developing and to cover the literature in a couple of sentences can miss the essence and serious works. (2) There is a restriction on the size of the paper, which does not allow both a classroom experience and a literature review.

In the present paper, we mention only those results which have strong connection with the detailed methods and tools. These statistical analyses proved that the classical tool-centred approaches (push production systems) do not develop students' computational thinking skills, there is no knowledge built up in long-term memory [Csernoch et al., 2022] [Csernoch et al., 2023]. Furthermore, we also found proof that our pull system with the Sprego programming approach is much more effective than the tool-centred methods [Csernoch, 2014] [Csernoch & Biró, 2015] [Csapó et al., 2019] [Csapó et al., 2020]. The content awareness, which is our other concern, is proved effective and efficient in the works of the TPCK research [Angeli & Valanides, 2015] [Kadijevich et al., 2013]. Based on these results, we ventured into the subject integration field and found that while students work on a table, they get familiar with the content also. They learn local places and events, and make acquaintance with further subjects, like Beaufort scale (content selected for the present study), barely known in continental countries.

Considering connection between knowledge built up in long-term memory and the role of repetition, we found notable results in second language acquisition research. It is said in language teaching that a word must be repeated over time which method is more effective than hearing a bunch of massed repetitions at one time. It is the same in computer sciences and informatics. While meeting the Beaufort scale in the printed course book only three times, in informatics classes, the word is mentioned so many times and the content is checked repeatedly that students learn both the name and the content. We have also found that to make students remember functions, they must be simple and be repeated in different contexts [Ghazi-Saidi & Ansaldo, 2017]. These reasons and findings explain the functionality of real world contents, repetition, and simplicity.

## 2. A CLASSROOM STUDY IN DIGITIZATION

### 2.1. The aims of the study

In the present paper, a classical paper-based problem is digitized, converted into a spreadsheet workbook within the framework of a classroom study [Stigler & Hiebert, 2009]. The research analysis reveals what skills can be developed by solving the course book tasks (Chapter 2.2), compared to the conversion process and the data management activities in spreadsheet with Sprego [Csernoch, 2014] [Csernoch & Biró, 2015] [Csapó et al., 2019] [Csapó et al., 2020].

Our ultimate goal is to prove that

- pull production system can be adapted in digital education and
- education production system can be as effective and efficient as it is in industry and service.

The target condition [Rother, 2010] [Liker, 2004] of this paper is to provide the essence of a classroom study which presents a method and its accompanying tools in a digital education pull production system. Further research is needed with novel partners (schools, companies, teachers, students, etc.) to statistically prove the effectiveness of the method.

In this education pull system, beyond developing teachers' and students' programming, data management, and spreadsheet knowledge, skills, and abilities, teaching and learning how to integrate sciences and school subjects and the normally (in push production systems) separately taught subjects of computer sciences have a great importance. With this method, both teachers and students learn the algorithms behind built-in, problem specific functions and spreadsheet features, and let go of software and/or software tools when problems require so. Two of the advantages are that (1) these fundamental algorithms and features can be used seamlessly (relying on schemata by activating fast thinking) [Kahneman, 2011] [Sweller et al., 2011] (2) and end-users have the background knowledge to select the most efficient tools to solve their problems (requiring slow thinking) [Kahneman, 2011] [Sweller et al., 2011] [Polya, 1945] [Smalley, 2018]. Furthermore, this system can reveal how classical, paper-based problems can be digitized which would serve a topic of data management, hardly covered yet.

However, we must keep in mind what Ohno said about TPS: "There is no real substance to that abstract mass we call 'the public'. We discovered that industry has to accept orders from each customer and make products that differ according to individual requirements." [Ohno, 1988]. In education, the public is the lot of students and end-users, and we must prepare dedicated classes and teaching-learning materials that suit their best interest. In these circumstances, all classes and students must be considered unique, but the shared results of classroom studies would help teachers in developing their teaching-learning materials.

## 2.2. Beaufort-scale

The task introduced in this paper arrives from a Grade 5 Sciences course book [Horváth et al., 2020], where a coloured Beaufort scale is presented (Figure 1), accompanied with three "classical" tasks.

- Use the table to determine the strength of the strongest wind you have ever encountered. Tell us what you experienced.
- What was the wind force described in the following short paragraph? Answer the question based on the Beaufort scale presented on page 110.
- Use the Beaufort scale to determine the approximate strength of the actual wind. If a suitable tool is available, measure the speed of the wind.

Figure 1. The original Hungarian paper-based (left) and the Excel version (right) of the Beaufort scale.

Figure 2. An English version of the Beaufort scale [RMetS, n.d.].

Several English Beaufort scale webpages are available on the internet (e.g., [Met Office, n.d.] [Edwards, n.d.] [US Department of Commerce NOAA, 2016] [RMetS, n.d.] [National Geographic, n.d.] [Encyclopædia Britannica, 2023] [Wikimedia Foundation, 2023]). In appearance and content, the closest to the course book version is [RMetS, n.d.] (Figure 2).

### 2.3. Concerning digitization

The primary aims of the present study are to provide the details of

- the conversion process from the original paper-based table to an Excel workbook,
- how the spreadsheet can support automated information retrieval, and
- how students can develop their skills and abilities in the TPCK process.

When the digitization of paper-based tasks or problems is in the focus, the main questions considering students' development are:

- what competencies and skills can be developed,
- what can be stored in long-term memory,
- how subject integration can support deep learning,
- how fast and slow thinking can be activated in the data retrieval process.

Further questions are how teachers are prepared to be able to develop students' content knowledge, their digital awareness, and their problem solving skills and abilities. In the following, a classroom study [Stigler & Hiebert, 2009] is detailed to present the digitization process of the selected content, the Beaufort scale in a Grade 8 group (13–14 years old students).

### 2.4. The description of the class

Students started to handle conversion processes in Grade 7 [Sebestyén et al., 2023] [Csernoch, 2019] (along with word processing), while spreadsheeting was launched at the beginning of Grade 8. Considering the content, at the beginning of the school year students confirmed that they never encountered Beaufort scale in advance to this project.

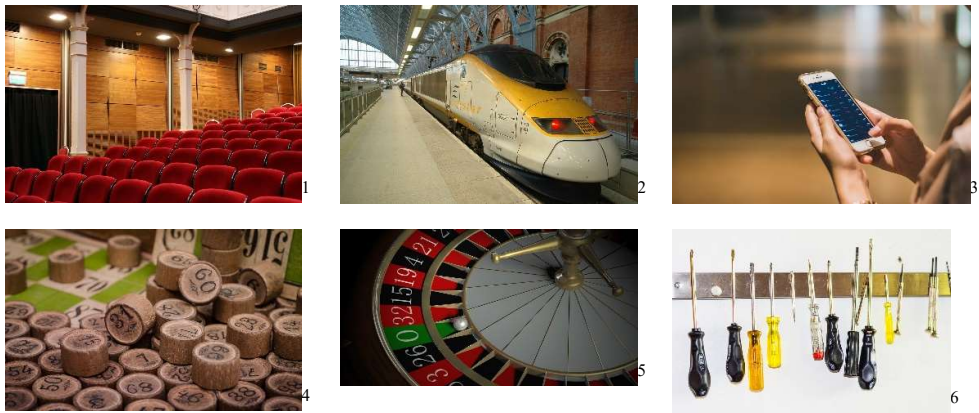

Figure 3. Real world examples for linear search.

At the beginning of the spreadsheet studies, a Hungarian Beaufort scale webpage [Magyar Turizmus Média Kft., 2019] was converted to a spreadsheet workbook to become familiar with the content, to repeat webtable→datatable conversion [Sebestyén et al., 2023] [Csernoch, 2019], and to become familiar with fundamental spreadsheet concepts. Following this introductory section, the algorithm of linear search was introduced with lots of examples from real world situations (e.g., finding the row and the seat in a theatre; finding the platform, the wagon, and the seat at a train station; searching in a mobile list; unordered items) (Figure 3).

After linear search, examples of binary search on ordered lists are presented in real world situations (e.g., playing the Find the number! game, searching in a dictionary, and finding the proper-size screwdriver) (Figure 4).

Following the introduction of both algorithms, students can decide which algorithm should be used depending on the input data. This section took three classes in October.

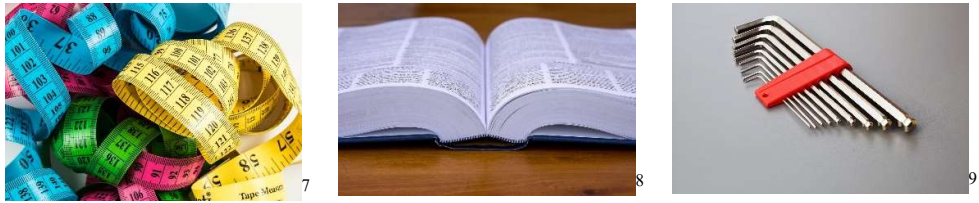

Figure 4. Real world examples for binary search.

The paper-based Beaufort scale was presented in the class in the second semester of the academic year. The students remembered the Beaufort scale and were delighted to meet a table which they were familiar with. The introduction of the content at the beginning of the year in a digital environment was enough to provide a firm background, based on which we were able to focus on the programming details. This section took six classes in January.

### 3. THE DATA MANIPULATION PROCESS

During the data manipulation process, the content of the webtable is converted into a 1st normal form datatable. The primary aims of the process are to make students

- become familiar with the content (again),
- recognize data fields, records, missing and redundant data,
- build the algorithm of the conversion process,
- find a suitable software environment or environments to carry out the planned process, and
- create an Excel-compatible format.

### 3.1. Analysis of the original table

The data management procedure starts with the analysis of the original table. In this step, the number of data fields and records should be identified, and the data types recognized. During this phase, students must keep in mind that the target format is a spreadsheet table. In this special table (originated from a printed course book), students should recognize that in the original first column three data are stored whose overall data type is string:

- data (content)
    - force,
    - the lower boundary of the speed,
    - the upper boundary of the speed,
- additional characters
    - parenthesis around the speeds,
    - n-dash between the lower and the upper boundary,
    - force and speeds are broken into two paragraphs.

Furthermore, in cells A2, A14, and C14 there is missing data (Figure 1, left panel).

- In A2, there is only one speed.
- In A14, there is only one speed with an additional < character.
- C14 is empty.

The redundancy in the first column should also be recognized. Teachers must find ways to convince students that there is no need to store both the lower and the upper boundaries of the speed. Recognizing and handling redundancy is a complex problem, and it is always the teachers' responsibility to introduce this concept, as early as possible [Hattie, 2012].

In general, the final Excel worksheet should contain 4 data fields, 1 row as header (field names) and 13 rows as data records. The data types in Columns A and B should be number, while in C and D string (Figure 1, right panel).

### 3.2. Conversion process: from PDF to DOCX

The conversion process starts with a file-save operation, where the page of the Beaufort scale is printed as a PDF file. This file can be opened in MS Word for further data handling (it could also be opened in Excel, however, for a beginner that task is too demanding). The left pane in Figure 5 shows the results of the conversion, and a compulsory first clean up phase which deletes data outside of the table.

The result is a tree-column table where some data are lost in the conversion process, but these minor errors can be corrected, along with the entry of the missing data (Figure 5, right).

| Fokozat¶ (km/óra) | Megnevezés | Szárazföldön |
|---|---|---|
| 0¶ (0) | szélcsend | a füst egyenesen száll fel |
| (2-6) | gyenge szellő | a füst gyengén ingadozik |
| (7-12) | könnyű szél | a fák levelei mozognak |
| (13-18) | gyenge szél | a vékony gallyak erősen mozognak |
| 4¶ (19-26) | mérsékelt szél | kisebb ágak is mozognak |
| 5¶ (27-35) | élénk szél | nagyobb ágak mozognak |
| 6¶ (36—44) | erős szél | nagyobb ágak erősen mozognak |
| 7¶ (45-54) | igen erős szél | vékony fatörzsek hajladoznak, gallyak törnek |
| 8¶ (55-65) | viharos szél | vastagabb fatörzsek hajladoznak, kisebb ágak törnek |
| (66-77) | vihar | kis fák kidőlnek, nagyobb ágak törnek |
| 10¶ (78-90) | erős vihar | épületekben, tetőkben nagyobb kár, kidőlt fák |
| 11¶ (91-104) | igen erős vihar | még súlyosabb pusztítás |
| 12¶ (105 | orkán | |

| Fokozat¶ (km/óra) | Megnevezés | Szárazföldön |
|---|---|---|
| 0¶ (0-1) | szélcsend | a füst egyenesen száll fel |
| 1¶ (2-6) | gyenge szellő | a füst gyengén ingadozik |
| 2¶ (7-12) | könnyű szél | a fák levelei mozognak |
| 3¶ (13-18) | gyenge szél | a vékony gallyak erősen mozognak |
| 4¶ (19-26) | mérsékelt szél | kisebb ágak is mozognak |
| 5¶ (27-35) | élénk szél | nagyobb ágak mozognak |
| 6¶ (36-44) | erős szél | nagyobb ágak erősen mozognak |
| 7¶ (45-54) | igen erős szél | vékony fatörzsek hajladoznak, gallyak törnek |
| 8¶ (55-65) | viharos szél | vastagabb fatörzsek hajladoznak, kisebb ágak törnek |
| 9¶ (66-77) | vihar | kis fák kidőlnek, nagyobb ágak törnek |
| 10¶ (78-90) | erős vihar | épületekben, tetőkben nagyobb kár, kidőlt fák |
| 11¶ (91-104) | igen erős vihar | még súlyosabb pusztítás |
| 12¶ (105-500) | orkán | a szél épületeket, tetőket rombol, súlyos pusztítást végez |

Figure 5. The converted Beaufort table opened in MS Word (left) and the corrected table (right).

### 3.3. Setting up data fields

After correcting and unifying the data in the first column (Figure 5, right panel), the separation of the three data – force, lower boundary and upper boundaiesy of speed – and the deleting of the unnecessary characters – Enter, parentheses, and dash – must be performed. These steps are required before the proper data fields are set up.

The listed modifications can be completed with replacement, where

- Enter and opening parenthesis are replaced with one Tab character (Figure 6, left),
- dash is replaced with one Tab character (Figure 6, middle),
- closing parenthesis is removed (replaced with nothing) (Figure 6, right).

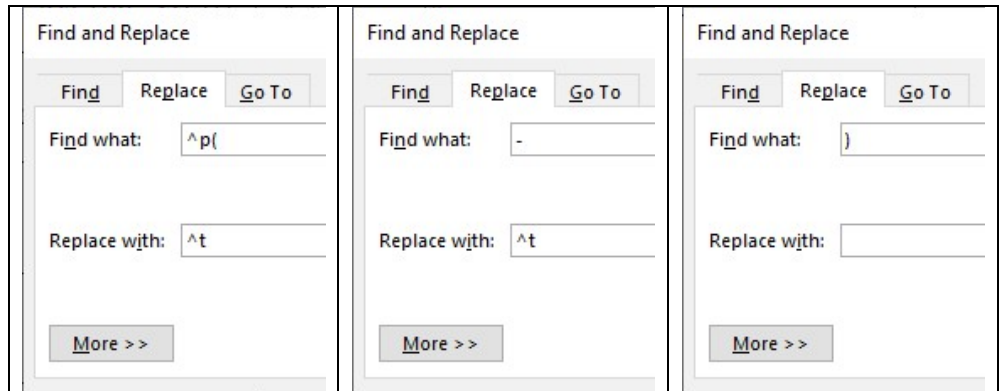

Figure 6. A series of replacements to separate the three data in the first column.

With the *Replace All* command the number of replacements are presented in the information window, where we can check whether the number is correct or not. These numbers are 14, 13, and 14, respectively. In the first and the third replacements, the 13 records and the field name, while in the second, only the records contained the searched string.

| Fokozat → seb1 → seb2 | Megnevezés | Szárazföldön |
|---|---|---|
| 0 → 0 → 1 | szélcsend | a füst egyenesen száll fel |
| 1 → 2 → 6 | gyenge szellő | a füst gyengén ingadozik |
| 2 → 7 → 12 | könnyű szél | a fák levelei mozognak |
| 3 → 13 → 18 | gyenge szél | a vékony gallyak erősen mozognak |
| 4 → 19 → 26 | mérsékelt szél | kisebb ágak is mozognak |
| 5 → 27 → 35 | élénk szél | nagyobb ágak mozognak |
| 6 → 36 → 44 | erős szél | nagyobb ágak erősen mozognak |
| 7 → 45 → 54 | igen erős szél | vékony fatörzsek hajladoznak, gallyak törnek |
| 8 → 55 → 65 | viharos szél | vastagabb fatörzsek hajladoznak, kisebb ágak törnek |
| 9 → 66 → 77 | vihar | kis fák kidőlnek, nagyobb ágak törnek |
| 10 → 78 → 90 | erős vihar | épületekben, tetőkben nagyobb kár, kidőlt fák |
| 11 → 91 → 104 | igen erős vihar | még súlyosabb pusztítás |
| 12 → 105 → 500 | orkán | a szél épületeket, tetőket rombol, súlyos pusztítást végez |

Figure 7. The data fields with mixed separators.

The result of the data separation process is a table where two column separators are: tabulators and vertical lines (Figure 7). In the field-setup process, the next step is the unification of separators. Furthermore, we can decide whether we delete the upper boundary of speed in Word or we can leave it to Excel. In the presented solution, the first option is performed (Figure 8). The computer cooking (CoCoo for short) sequence is presented in CoCoo 1.

```
Fokozat →  seb1 → Megnevezés      → Szárazföldön¶
0       →  0    → szélcsend       → a füst egyenesen száll fel¶
1       →  2    → gyenge szellő   → a füst gyengén ingadozik¶
2       →  7    → könnyű szél     → a fák levelei mozognak¶
3       →  13   → gyenge szél     → a vékony gallyak erősen mozognak¶
4       →  19   → mérsékelt szél  → kisebb ágak is mozognak¶
5       →  27   → élénk szél      → nagyobb ágak mozognak¶
6       →  36   → erős szél       → nagyobb ágak erősen mozognak¶
7       →  45   → igen erős szél  → vékony fatörzsek hajladoznak, gallyak törnek¶
8       →  55   → viharos szél    → vastagabb fatörzsek hajladoznak, kisebb ágak törnek¶
9       →  66   → vihar           → kis fák kidőlnek, nagyobb ágak törnek¶
10      →  78   → erős vihar      → épületekben, tetőkben nagyobb kár, kidőlt fák¶
11      →  91   → igen erős vihar → még súlyosabb pusztítás¶
12      →  105  → orkán           → a szél épületeket, tetőket rombol, súlyos pusztítást végez¶
```

Figure 8. The data fields with a unified separator character, the tabulator.

> the table with the mixed separators (Figure 7) is converted into text with Tabulator separator (*Convert to Text*) → the text is converted into a table of 5 columns (*Convert Text to Table*) → the column of the upper boundary of speed is deleted → the table is converted back to text (*Convert to Text*) (Figure 8).

CoCoo 1. The major steps of the unification of the field-separator characters.

### 3.4. Creating Excel workbook: from DOCX to XLSX

At this phase, the table is ready to be converted into an Excel workbook. One of the simplest ways to create an Excel workbook from a Word document is to use an interim file format, which is text file (*Plain Text*) with Unicode (UTF-8) encoding. The algorithm is presented in CoCoo 2.

> the table is converted into text with a unified separator (e.g., tabulator) → the Word document is converted into a text file → the text file is opened in or imported into Excel → the text file is converted into an Excel workbook

CoCoo 2. The conversion process from Word document to Excel workbook.

### 3.5. Mapping

As a warm-up activity, a diagram was created, based on the data converted into an Excel workbook. The category labels are forces, while the values are the lower boundary of the speed (speed for short) (Figure 9).

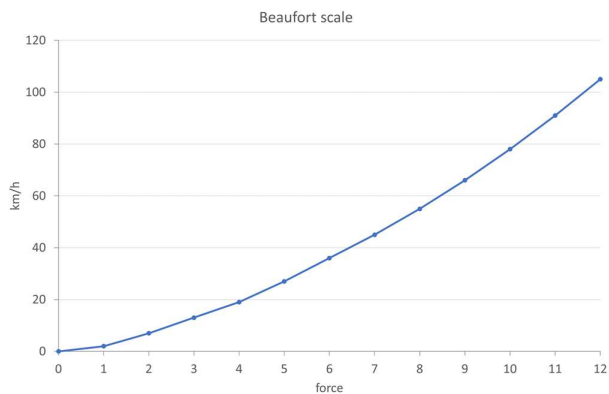

Figure 9. Beaufort scale diagram.

## 4. COLOURING

The next step of the preparation process is the colouring of the Excel table (Figure 10, Columns A–D).

Figure 10. The coloured Beaufort scale (Columns A–D) with the RGB codes (Columns E–G) and the picture of the original table (Column I). (Specifications are shortened in Column D.)

To complete the colouring task, a picture was created from the original table of the course book (Figure 10, Column I). Based on the picture, the RGB codes of the colours (Figure 10, Columns E–G) were collected with a picker tool, which can be carried out either in a professional image manipulation program (usually we use GIMP, since it is available in our schools) or in a simple PowerPoint.

## 5. INFORMATION RETRIEVAL

### 5.1. Linear and binary searches

In the simplest retrieval tasks, we can provide the force or the speed as input (Figure 11, Cells F3 and F7, respectively) and write out the other three data. The difference between the two inputs is that the value of force can only be an integer from 0 to 12 (listed in the vector), while the speed can be any number greater or equal to 0. Consequently, the algorithm to find them in the corresponding vector is different (Force or Speed, respectively). In the case of speed, binary search must be used, while with force both linear and binary search can be applied. However, for students, linear search is more understandable, consequently it is recommended to use this simpler algorithm.

Figure 11. Force (Cell F3) and speed (Cell F8) input values are entered by the user, and the other three values are calculated based on the Beaufort table.

In both cases, the algorithm is provided in CoCoo 3.

finding the position of the input value in the corresponding vector (Formulas (1) and (3)) → writing out the value from another vector at the same position (Formulas (2) and (4))

CoCoo 3. The steps of the searching process.

| =MATCH(F3,A2:A14,0) | (1) |
| =INDEX(B2:B14,MATCH(F3,A2:A14,0)) | (2) |

Providing force as input, the linear search algorithm first finds the position of the given force in the corresponding vector (Force). After that, the speed is written out in the second step.

With the same match_type value (0), both the description and specification can be written out (Figure 11, Cells H3 and I3).

| =MATCH(F8,B2:B14) | (3) |
| =INDEX(A2:A14,MATCH(F8,B2:B14)) | (4) |

Providing speed as input, Formula (4) writes out the corresponding force (Figure 11, Cell G8). With the MATCH() function, using match_type=1 value (default value) a binary search is launched. With the same procedure, both the description and specification can be written out (Figure 11, Cells H8 and I8).

As the last step of these practices, complete sentences can be created with the combination of formulas and string constants (Figure 12).

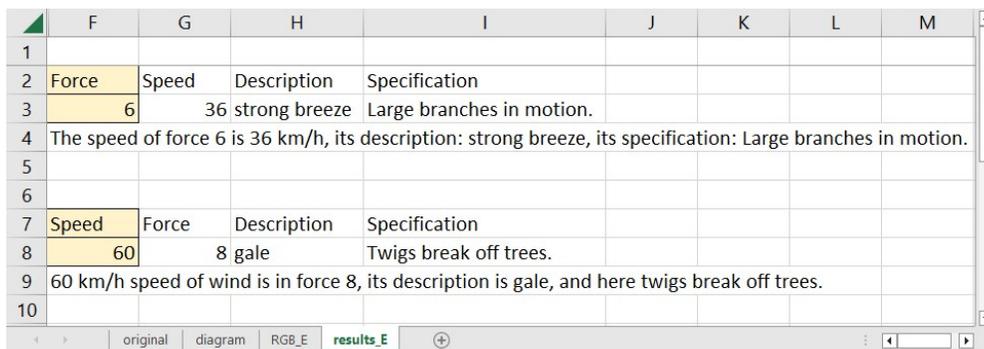

Figure 12. Complete sentences in Cells F4 and F9 with the given input values in Cells F3 and F8.

### 5.2. Coloured sentences

All the output values can be coloured with the original colour of the corresponding force. To generate such outputs a conditional formatting is used with RGB codes gained from the original table (Figure 10).

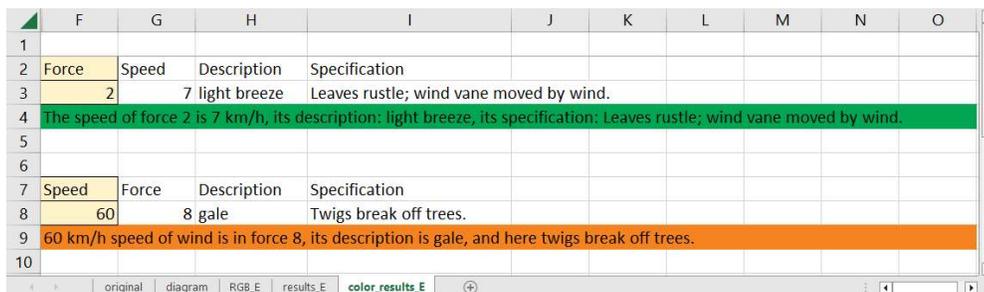

Figure 13. Complete sentences with the corresponding colour of force.

### 5.3. Drop-down lists

In the present paper details of drop-down lists created to the description vector is presented. However, in a similar way drop-down lists can be created for the specifications and even for the forces.

To select descriptions, drop-down lists are more convenient than typing these long strings. Two types of objects can be inserted into the sheets:

- Form Controls (Figure 14, Cell G3),
- ActiveX Controls (Figure 14, Cell G10).

In both cases, two parameters of the object must be set up:

- Input range (ListFillRange), the values (items) to appear in the list (Figure 10, Range C2:C14),
- Linked cell (LinkedCell), the cell displaying the output (Figure 14, Cells F3 and F10).

The difference between the two types of object is the output:

- Form Control returns the position of the item in the vector (Figure 14, Cell F3) (the index of the selected item),
- ActiveX Control returns the selected item (Figure 14, Cell F10).

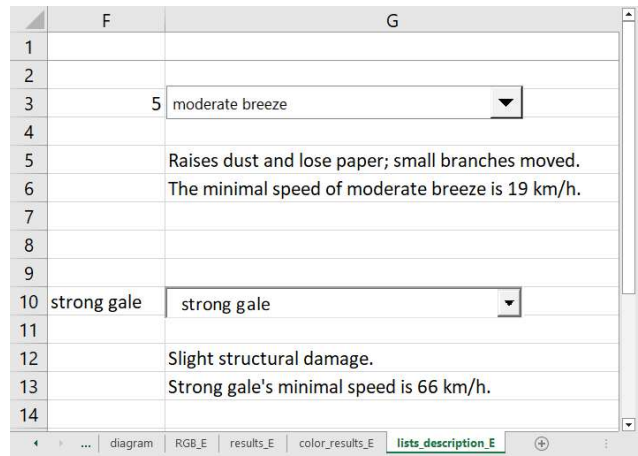

Figure 14. Drop-down lists for selecting specification.

Based on the selected description, the other data of the records can be presented in the table. To write out the specification for the item selected from a Form Control a simple INDEX() function serves our purposes (Figure 14, Cell G5) (Equation (5)).

| =INDEX(D2:D14,F3) | (5) |
|---|---|

To create a complete sentence with the selected description and the corresponding speed, two INDEX() functions (Formulas (5) and (6)) can be called with the additional string constants (Figure 14, Cell G6).

| =INDEX(B2:B14,F3) | (6) |
|---|---|

In the case of ActiveX Control, to write out the specification (Figure 14, Cell G12) the 2-level function of INDEX(MATCH()) can be called (Formula (7)).

| =INDEX(D2:D14,MATCH(F10,C2:C14,0)) | (7) |
|---|---|

To create a complete sentence with the selected description and another data from the record (Figure 14, Cell G13) a reference to the output cell (LinkedCell) (Formula (8)) and the 2-level INDEX(MATCH()) can be called (Formula (7)), with the additional string constants.

| =F10 | (8) |
|---|---|

Capitalizing the first letter of the sentence requires some text handling, whose algorithm is presented in CoCoo 4.

> separating the first letter of the description → changing the lower-case letter to capital → separating all the characters except the first one of the description from the right side → concatenating the two strings (Formula (9)) (Figure 14, Cell G13)

CoCoo 4. The algorithm of capitalizing the first letter of the sentence.

| =UPPER(LEFT(F10))&RIGHT(F10,LEN(F10)-1) | (9) |
|---|---|

The outputs based on a selection from a drop-down list can also be coloured, similar to sentences presented in Figure 13.

## 6. SUMMARY

### 6.1. Manual search in the table

In the original course book tasks (Chapter 2.2), students were required to carry out three manual searches in the presented table to provide two times the speed and once the force of the wind experienced or described in the book. The tasks would take five minutes. Beyond completing these manual searches in the Beaufort scale, students never encounter the table again.

### 6.2. Digitization of the table

In the digitization process of the table, students carry out several different activities, and learn how to

- set up a data table (1st normal form), fields, records, column separators,
- analyse and correct data,
- enter the escape sequences of non-printable characters in Word,
- convert text to table and table to text in Word,
- handle different data types (in this task: webpage, pdf, Word document, text file, Excel workbook, image),
- copy and name worksheet, freeze panes,
- insert and resize image on a worksheet,
- read RGB codes of an image,
- create and format diagram,
- enter input data with methods which are more secure and faster than typing (dropdown lists),
- create multi-level functions,
- handle n-ary functions,
- use the algorithm of linear and binary searches,
- create flexible natural language sentences with variables and formulas,
- set up conditional formatting,
- carry out debugging.

The digitization of the course book table was introduced with the conversion of a Beaufort webpage. The two processes took nine classes together. During these classes, students had the opportunity to get familiar both with the name and the content of the table and learn and practice the basis of functional programming in spreadsheet.

### 6.3. Improvement of spreadsheet quality

The essence of a digital education pull production system is that only those fundamental knowledge pieces and tools are introduced and taught which are needed to solve the presented problems, matching the characteristics of the students (end-users). With this method in the first learning period, slow thinking is activated and schemata are built up (standardization) [Sweller et al., 2011] [Kahneman, 2011] [Polya, 1945]. Later, fast thinking can activate schemata to fast and reliable problem solving. Considering these roles of fast and slow thinking in spreadsheet education, we provide some examples connected to already existing features.

E.g., pivot table is an excellent tool. However, it is a horror for those who do not understand the algorithms behind it. To avoid this discrepancy, in Sprego first, we teach and thoroughly practice conditional calculations along with conditional formatting then we can try pivot tables. On the same background with conditional operations, we can eliminate the specified conditional and database functions with all their restrictions.

E.g., the table in the table feature is one of the newest and useless features in education. If end-users learn how to create 1st normal form table, there is no need for the *Format as table* command. Furthermore, knowledge connected to normal forms can be transferred to database management and programming.

E.g., if students learn fundamental algorithms in spreadsheet environment, later they can transfer this knowledge to 'serious' programming.

It is said at Toyota that "Take Good Care of Old Equipment. One source of eliminating waste is not to buy new computers (digital devices) just because they appeared on the market. … A machine's value is not determined by its years of service or its age. It is determined by the earning power it still remains." [Ohno, 1988]. This saying is true with computer sciences and informatics; just because a new hardware, software, or a new feature in a software appears, there is no need to rush head over heels to get it. Teaching fundamental algorithms and general-purpose functions allow us to create program- and version-independent spreadsheets that can be applied in any interface, while it is not true with fancy new features.

### 7. DISCUSSION

Beyond the traditional sense of subject integration, novel concepts, including TPCK [Angeli & Valanides, 2015] [Kadijevich et al., 2013] and digital problem solving, slowly though, but are advancing and reaching education. Within this framework, another subject integration can be introduced, namely the connection and knowledge-transfer between the various subjects of informatics (computer sciences). The main difference between the two types of subject integration is that this latter often raises debates and is seldom accepted in spite of its unquestionable values.

In the present paper, an example of digital subject integration is detailed. The subject is the Beaufort scale of winds – adapted from a traditional Grade 5 Sciences course book. The original paper-based table is converted into an Excel workbook, and in this GUI simplified functional programming is performed (exaptation, defined in [Hatamleh & Tilesch, 2020]). The paper details the data analysis and the conversion process, and provides examples of data retrieval methods in spreadsheets, along with the algorithms concerned.

It is found that the introduction of the selected table in a digital environment helps students to deepen their content knowledge in sciences and offer a real-world problem for programming, data management, and information retrieval. During the data analysis and handling process, especially in the discussion section, students must be aware of both the concept and the content of the table. Furthermore, the conversion process requires skills that strengthen the connection between different data management approaches and tools. The combination of these skills deepens students' knowledge in both sciences and informatics, which is one of the aims of education.

In general, the presented method is part of a huge project which ultimate goal is to find ways to set up a digital education pull production system. The TPS pull production system [Ohno, 1988] [Krafcik, 1988] [Womack & Jones, 2003] [Modig & Åhlström, 2018] is proved efficient in industry and service, so we are convinced that the low-growth period of digital education can benefit from such changes.

**SOURCES**

---

[1] https://pixabay.com/images/id-1727890/ 12.30pm  23/05/27
[2] https://pixabay.com/images/id-911760/ 12.30pm  23/05/27
[3] https://pixabay.com/images/id-7304257/ 12.30pm  23/05/27
[4] https://pixabay.com/images/id-1044718/ 12.30pm  23/05/27
[5] https://pixabay.com/images/id-2001129/ 12.30pm  23/05/27
[6] https://cdn.pixabay.com/photo/2014/11/09/15/35/the-screwdriver-523783_1280.jpg 12.30pm  23/05/27
[7] https://pixabay.com/images/id-3509492/ 12.30pm  23/05/27
[8] https://pixabay.com/images/id-1798/ 12.30pm  23/05/27
[9] https://pixabay.com/images/id-2134715/ 12.30pm  23/05/27